\def\be{\begin{equation}}
\def\te{\end{equation}}

\documentstyle[12pt]{article}
\begin{document}

\begin{center}
{\bf QUANTUM AND THERMAL FLUCTUATIONS, UNCERTAINTY PRINCIPLE,
DECOHERENCE AND CLASSICALITY}
\footnote{\it Invited Talk delivered by B. L. Hu
at the Third Workshop on Quantum Nonintegrability,
Drexel University, Philadelphia, May, 1992.
To appear in {\rm Quantum Dynamics of Chaotic Systems},
edited by J. M. Yuan, D. H. Feng and G. M. Zaslavsky
(Gordon and Breach, Langhorne 1993)}
\vskip 1cm
{\bf B. L. Hu }\\
{\sl Department of Physics, University of Maryland, College Park, MD 20742,
USA}
\vskip .5cm
{\bf Yuhong Zhang}\\
{\sl Biophysics Lab, CBER, Food and Drug Adminstration,
8800 Rockville Pike, Bethesda, MD 20982, USA}
\vskip .5cm
(preprint umdpp 93-58)\\
\vskip .5cm
\end{center}

\begin{abstract}
 We scrutize the commonly used criteria for classicality and examine
their underlying issues. The two major issues we address here
are that of decoherence and fluctuations. We borrow
the insights gained in the study of the semiclassical limit of
quantum cosmology to discuss the three criteria of classicality for a
quantum closed system: adiabaticity, correlation and decoherence.
We then use the Brownian model as a paradigm of quantum
open systems to
discuss the relation of quantum and thermal fluctuations
and their role in the transition from quantum to classical.
We derive the uncertainty relation at finite temperature.
We study how the fluctuations of
a quantum system evolve after it is brought in contact with a heat bath
and analyse the decoherence and relaxation processes.
{}From the effect of fluctuations on decoherence we show
the relation between these two sets of criteria of classicality. Finally,
we briefly comment on the issue of nonintegrability in quantum open systems.
\end{abstract}
%\noindent $^{\ast}$: bitnet addresses: hu@umdhep, zhang@umdhep

\newpage

\section{Criteria for Classicality}

A quick sampling of text books of quantum mechanics and statistical mechanics
reveals a variety of seemingly simple and straightforward
criteria and  conditions
for classicality. For example, one can loosely associate:\\
1) $\hbar \rightarrow 0$\\
2) WKB approximation, which "gives the semiclassical limit"\\
3) Wigner function, which "behaves like the classical distribution function"\\
4) high temperature limit: "thermal=classical"\\
5) Uncertainty Principle: a system "becomes classical" when this is no longer
obeyed\\
6) coherent states:  the `closest' to the classical\\
7) systems with large quantum number $n \rightarrow \infty$
(correspondance principle)\\
8) systems with large number of components $1/N \rightarrow 0$

Each of these conditions contains only some partial truth and when taken on
face
value can be very misleading. To understand the meaning of classicality
it is important to examine the exact meaning of these criteria
and the conditions of their validity.\\

We can divide the above conditions into four groups, according to the
different issues behind these criteria:\\
a) quantum interference in 1) to 3),\\
b) quantum and thermal fluctuations in 1),4) and 5),\\
c) choice of special quantum states, and\\
d) meaning of the large n and N limit.

We will discuss only the first two groups of issues, using lessons we learned
from quantum cosmology and the paradigm of quantum open systems. We will
show that many of these criteria hold only under special conditions. They
can approximately define the classical limit only when taken together in
specific ways. We will
also show the relation of the first two groups of criteria. Specifically,
for issue a), decoherence is needed
for the WKB Wigner function to show
a peak indicating the correlation between the physical
variables and their canonical conjugates which defines a
classical trajectory in phase space.
This clarifies the loose connection of WKB, Wigner function and classicality.
For issue b), the time for thermal fluctuations to overtake quantum
fluctuations
is also the time of decoherence. But a decohered system is not necessarily
classical. There is a quantum statistical regime in between.
One can derive an uncertainty principle
for a quantum open system at finite temperature which interpolates between
the (zero temperature) quantum Heisenberg relation and the high temperature
result of classical statistical mechanics. This clarifies the sometimes
vague notions of quantum, thermal and classical.
The first set of issues was discussed in the context of quantum
cosmology by the authors of \cite{decQC}. We will only state the results
without
derivation. The second set of issues was clarified in a recent paper of ours
\cite{HuZhaUncer}, which the present report is based on.
We will discuss these issues in the following sections. We end with a few brief
comments on the main theme of this conference with the viewpoint espoused here,
i.e., how one should
perceive  quantum and classical chaos, when quantum and classical physics
are taken to be the physics of a closed versus an open system.

\section{Semiclassical Limit: Adiabaticity, Correlation, and Decoherence}

Our interest in the problem of quantum to classical transition stemmed from
related problems in quantum cosmology. How did the classical world as we
observe it today evolve from a wave function or density matrix
of the universe \cite{WaveUniv}? An important transition point is the Planck
time ($10^{-43} sec$).
The belief is that only after this time can
the world be described by semiclassical gravity, a theory
where the spacetime is classical but the matter field is quantized.
The issue of transition from quantum to classical spacetimes
is complicated here by
two additional factors unique to quantum cosmology (as an example of a
quantum closed system described as a parametrized theory): there is no
`external' classical observer to carry out a measurement,
and there is no `explicit' time variable to define the evolution \cite{timeQC}.
We shall try to comply with these conditions in this part of our discussion.
It is easier to relax these conditions when there is a well-defined
time, and when there is a reasonable system-environment separation.
After five years of search and research, one can
say that at least three conditions are necessary \cite{PazSin1,PazSin2,HalGR13}
%Using the minisuperspace models  and analyzing the form of the
%reduced WKB Wigner function Sinha and Paz (1991) derived three conditions
for the apperance of  classical behavior in spacetime:
i) the adiabaticity condition,  which ensures that the leading terms of
%restricts  how much higher-order WKB solutions can contribute.
a WKB expansion of the wave function dominate;
ii) the strong correlation condition , by  requiring a sharp peak in the
reduced WKB Wigner function; and
iii) the strong decoherence condition, by requiring the close diagonalization
of the reduced density matrix.
In this section we follow the summary in \cite{SinhaPhD}:

\subsection{WKB}

The first condition to see the emergence of classical spacetime
is that certain regions of superspace (the space of all three-geometries)
has to admit oscillatory
solutions to the wave equation, which in quantum cosmology
is known as the Wheeler DeWitt equation.
For a universe consisting of a gravitational sector described by coordinates
$r^a$ of the minisuperspace variables (for the Robertson-Walker universe,
$r$ is the scale factor $a$) and a matter-field sector
whose coordinates are the amplitude functions of the normal modes $f_k$ of,
say, a scalar field $\Phi$,
the Wheeler-DeWitt equation for the wave function
of the universe $\Psi(r^a, f_k)$ can
be written in the general form
\begin{equation}
\left[{1\over {2 M}} {\nabla}^2+M V(r^a)-H_{\Phi}(r^a,\Phi) \right] \Psi(r^a)
=0
\end{equation}
where $M$ is a large parameter (in units of Planck mass squared),
${\nabla}^2$ is the Laplacian
and $V(r)$ is a potential function on the minisuperspace
(which can contain contributions from spatial curvature, the cosmological
constant and classical matter), and
$H_{\Phi}= \Sigma_k H_k$ is the sum of the hamiltonians for each mode of the
matter field.
[We can denote these variables schematically as $ (r, f)$.
Later, in the discussion of quantum open systems, we can
think of them as the system and environment variables.]
Classical limit exists only in those regions of superspace
where one can write the wave function in the WKB form:
\begin{equation}
\Psi(r,f) = e^{iMS(r)}C(r)\chi(r,f).
\end{equation}

The existence of a large parameter $M$ in the theory
which measures the discrepancy between
the gravitational and the matter sector makes it possible to carry
out an expansion (Born-Oppenheimer approximation) of
all the functions $S, C, \chi$ in powers of $M^{-1}$, i. e.,
\footnote{In the case of pure gravity,
a  large $M$ expansion (Born-Oppenheimer) is equivalent to a small $\hbar$
expansion. However, these two expansions
are not equivalent when matter fields are included.}
\begin{equation}
S     =  S_{0}(r) + M^{-1}S_{1}(r) \ldots
\end{equation}
%\begin{eqnarray}
%C    & = & C_{0}(a) + M^{-1}C_{1}(a) \ldots \\
%\chi & = & \chi_{0}(a) + M^{-1}\chi_{1}(a) \ldots
%\end{eqnarray}
One obtains a set of interlinked equations for these functions of different
orde
   rs.
In particular, the order $M^0$ terms yield an equation for $S_0$ which is the
Hamilton-Jacobi equation for the eikonal function:
\begin{equation}
{1\over 2} {\left({dS_{0}\over dr}\right)^2} - V(r) = 0
\end{equation}
while the order $M^{-1}$ terms yield an equation for $S_1$ which
gives the WKB prefactor.
%\begin{equation}
%i{dS_{0}\over da}{\partial\chi_0\over \partial a} = h\chi_0
%\end{equation}
%\noindent if
%\begin{equation}
%2C_0 S'_0 + S''_0 = 0.
%\end{equation}
%(This equation allows us to write $C_0$ in terms of $S_0$
%but one should realize that there is an ambiguity
%in splitting the zero order contribution between
%$\chi_0$ and $C_0$. The usual choice for $C_0$
%is to define it in the same way as in the case in which the
%the system (the gravitational sector) is isolated, i.e. $h=0$.)

In quantum gravity time is an emergent quantity believed to be well-defined
only when the classical spacetime takes shape. In the WKB regime
one can define such a time by
\begin{equation}
{d\over dt} = {dS_0\over dr}{d\over dr}
\end{equation}
In terms of this WKB time, the equation for the lowest order prefactor
assumes the form of a Schr\"odinger equation
\begin{equation}
i{d\chi_0 \over dt}  =  H_{\Phi} \chi_0
\end{equation}
Thus the WKB approximation renders the wave mechanics of the
universe into a form like ordinary quantum mechanics.

\subsection{Correlations}

To make predictions in a quantum closed system like the universe,
one needs to develop a viewpoint beyond the Copenhagen interpretation of
quantum mechanics which does not rely on the
existence of a classical external observer or apparatus.
It has been proposed \cite{peak} that one can
regard a strong peak in the wave function or a distribution
constructed from it as predicting some correlations among the variables
in the support of the peak. (In the absence of a strong peak no
predictions can be made.) This proposal also avoids the
use of probabilities conserved with respect to an external time.
(See, however, \cite{Isham})

Following this interpretive scheme one can propose the following criterion:
a system can be
regarded as behaving classically when the wave function predicts the
existence of correlations between a physical variable and its canonical
conjugate such as coordinate and momentum.
These correlations should be such that
the classical equations of motion are satisfied.

A more direct way of implementing the correlation criterion \cite{correlation}
is to look for peaks in a quantum phase space distribution function about
such correlations.
The closest quantum mechanical analog of a classical phase space
probability distribution is the Wigner function \cite{Wigner}
%given by
%\be
%W(X, p) = {1\over \hbar} \int dy~ e^{{-2ipy} \over \hbar}
%\rho(X-{y \over 2}, X+{y \over 2})
%\te
%where $\rho (x_1 , x_2)$ is the density matrix in the coordinate
%representation, written in terms
%of the `average' and `difference' variables $ X= {x_1 + x_2 \over 2}$
%and $y = x_1 - x_2 $ , and $p$ is the momentum.  It is
which is essentially the Fourier transform of the density matrix
[see, e.g., Eq.(5.3)].
\footnote{The Wigner function can however qualify
only as a quasiprobability distribution
because in general $W(X,p)$ is not positive definite
(it essentially reflects the fact that
in quantum mechanics $X$ and $p$ are not simultaneously measurable).
In spite of this property it can be used to obtain expectation values
in quantum mechanics by integrating over phase space and in this
respect presents a formulation of quantum mechanics exactly equivalent
to that given by wave functions or density matrices.
To interpret the Wigner function
as a distribution function , we will only use wave functions
for which it is indeed positive definite. For a more detailed discussion
on the subtleties of Wigner functions in quantum cosmology,
see \cite{HabLaf}.}
%Thus a peak in the
%Wigner function about a trajectory in phase space gives the desired
%momentum-coordinate correlations directly.
%
Initially it was claimed
that a calculation of the Wigner function for a WKB wave function gave
a delta function peak around the correlations.
Habib \cite{Hab} and  Anderson \cite{And}
using  earlier results of Berry \cite{Ber} and Heller \cite{Hel}
showed that this claim
was incorrect.  In fact the WKB Wigner function exhibits no peak around
classical trajectories and is highly oscillatory. To seek a peak
one needs to decohere the Wigner function.
This brings in the next criterion.

\subsection{Decoherence}

In quantum physics one must
assign a complex probability amplitude to each history. When combined
with the superposition principle this implies the existence of quantum
interference effects amongst alternative histories.
These effects are nevertheless not seen at the classical level.
Classical behavior in a system therefore requires some mechanism for the
destruction of quantum interference. This property is commonly referred
to as decoherence.
As  keys to understanding the emergence of classical behavior in quantum
closed systems Griffiths \cite{Griffiths} and Omnes \cite{Omnes}
have proposed the concept of consistent  histories, and Gell-Mann and Hartle
\cite{GelHar1} have proposed the use of decoherence functional as a measure
of the interference between histories.
One can also use the concept of environment-induced
decoherence in quantum open systems
to address these issues. An open system is one
which interacts with an environment whose information is not fully preserved.
In  coarse-graining the environment one introduces certain statistical
measures. Its averaged effect on the system appear as
dissipation and decoherence in the system dynamics.
%There, the Feynman-Vernon
%influence functional is a convenient quantity containing these information.
Decoherence was also discussed in the context of measurement
theory in quantum mechanics \cite{Zurek,JooZeh,WheZur}.
The connection between the density matrix approach and the
decoherence functional approach is a current topic of investigation.
\footnote{In the context of quantum cosmology, even though in theory the
universe is often regarded as containing  everything and thus
a closed system with no `external' environment, in practice our
observations are often restricted to a limited number of physical variables,
such as the homogeneous modes of cosmological perturbations, the low
energy limit of spacetime and field excitations, the causal
domain within the particle horizon, etc. In this sense, only these `relevant'
variables  constitute the physical `system', and the rest of
the universe are the irrelevant variables,  serving as an `environment'
to one's system of interest. This is how the basic paradigm of non-equilibrium
statistical mechanics becomes relevant to issues in quantum cosmology
\cite{HuTsukuba}.}

Therefore, apart from the existence of correlations,
a second necessary requirement for classical behavior in certain variables
is the  lack of interference between alternative histories of
those degrees of freedom implied by decoherence. Indeed these processes
are interconnected \cite{HuTsukuba}.
%Decoherence is needed to damp away the oscillatory
%behavior in the  Wigner function constructed from the WKB wavefunctions.
Incorporating the effect of the environment by working with the corresponding
reduced Wigner function rather than the pure state Wigner function,
Habib and Laflamme \cite{HabLaf} and Paz and Sinha \cite{PazSin1}
showed that such a peak indeed appears in the decohered WKB Wigner function
which follows the classical trajectory.
In addition, the environment will inevitably
have a backreaction on the system (just as in the case of Brownian
motion the environment produces a systematic damping in the motion
of the Brownian particle. See  \cite{FeyVer,CalLeg83,UnrZur,Gra,HPZ1,HPZ2}).
Paz and Sinha \cite{PazSin2} showed that under the adiabaticity condition
this backreaction indeed gives rise to the expected semiclassical Einstein
equations \cite{CalHu87}.

This is the story about how one obtains the semiclassical limit in
quantum cosmology. The many subtleties in how these criteria
are reached and applied constrast strongly with the simplistic
conditions listed in 1) to 3). One has to be very careful in the
interpretation  and application of these conditions.
The outstanding issue there is decoherence.
Let us now turn to the second set of issues,
concerning the role of quantum and thermal fluctuations in the transition from
quantum to classical. We will
derive the uncertainty principle at a finite temperature and use it to examine
these effects. We will see that the fluctuations are indeed what is responsible
for the decoherence process.
%Therefore the imposition of uncertainty principle
%as in condition 5)  under normal circumstances (e.g., ohmic environment
%at high temperature) gives the equivalent results as the decoherence
%%requiremen
   t.

\section{Quantum, Thermal and Classical}
\setcounter{equation}{0}

The usual definition or demarkation
of quantum, classical and thermal regimes are not always very precise.
Oftentimes one hears the vague statement (condition 4)
that high temperature regime gives
the classical limit.
%But how high is high, and at how low a temperature will
%the system behave quantum mechanically?
How does thermal properties enter into the criteria of classicality.
In particular, how do they
relate to the first set of criteria discussed above, e.g., decoherence?
What is the role of thermal fluctuations
in the establishment of classical behavior in a quantum system?
%We will point out with the aid of the results obtained here
%some existing confusions and make their meanings more precise.
%The relation
%between quantum and thermal fluctuations has been studied previously via
%%thermo
%field dynamics [8] under equilibrium and stationary conditions. We don't want
%to be restricted as such.
We will explore this issue using the model of quantum Brownian motion
as a paradigm of a quantum open system.
%Theoretically the only basic law we rely on is quantum mechanics.
We view thermal fluctuations not as the activity of some additional
`fundamental' field (such as is assumed in, e.g., thermal field dynamics
\cite{tfd} ), but simply as statistical variations of the physical
variables of a quantum system interacting with a stochastic source
associated with the environmental variables,
the exact microdynamics of the system and the environment obeying the laws
of quantum mechanics.
%The problem under study can thus be stated equivalently
%as finding the uncertainty relation in an open quantum system.
%On the third issue, in loose terms, one often identifies the high temperature
%regime of a system as the classical domain. On the one hand, one often regards
%the regime when thermal fluctuations begin to surpass quantum fluctuations
%as the transition point from quantum to classical. On the other hand, from the
%wave picture of quantum mechanics we know that a necessary condition for a
%system to behave classically is that the interference terms in its wave
%function have to vanish,
%
%reduced density matrix to become diagonal
%
%so that probability can be assigned to classical events [9]
%
%in the decoherent history viewpoint, for the decoherent functional to assume
%a diagonal form
%
%or that classical decoherent histories can be well-defined [10]. This is
%known as
%the decoherence process. Is there any relation between these two criteria of
%classicality?
We show that
under the conditions studied, the first group (decoherence) and second
group (fluctuations) of criteria give equivalent results.
The time the quantum system decoheres is also the time
when thermal fluctuation overtakes quantum fluctuations [see Eq. (27)].
However, we warn that the post-decoherence regime
%However we issue a warning here that this regime
should really not be called classical,
as is customary in many quantum to classical transition studies. In fact,
after the decoherence time, the system is aptly described by {\it quantum}
statistical mechanics (QSM), indeed, {\it non-equilibrium QSM}, not classical.
%only the first postulate of quantum statistical mechanics (QSM) is satisfied,
Only after the relaxation time
%when the second postulate is satisfied,
can one use {\it equilibrium QSM}.
Classical has still a long way to go---
%It is well-known that quantum statistical effects can be important at very
%%high
%temperatures (e.g., Fermi temperature for metals). This is due to exchange
%interactions of identical particles, a distinctly quantum effect.
only when the fermions and bosons in the system can be approximated
as distinguishable particles, usually at high temperatures when the Fermi-Dirac
or Bose-Einstein statistics approaches the Maxwell-Boltzmann statistics, can
the system be rightfully called {\it classical}. In this regard {\it quantum}
carries two meanings, one refers to the interference effect and the other
refers
to spin-statistics effect.

%The usage of the word classical in many decoherence studies
%is for practical considerations incorrect.

A clean-cut problem where these issues manifest clearly is the uncertainty
principle at finite temperature.
We learn from quantum mechanics that a lower bound exists in
the product of the variances of pairs of noncommutative observables.
Taking the coordinate $x$ and momentum $p$ as examples, the Heisenberg
uncertainty principle states that, with $(\Delta x)^2=~<x^2>-~<x>^2$,
the uncertainty function
\be
U_0 ^{QM} =(\Delta x)^2(\Delta p)^2 \geq {\hbar^2 \over 4}
{}~~~(T=0,~~ quantum~ mechanics).
%                                                              \eqno(1)
\te
%The existence of quantum fluctuations is a verified basic physical
%phenomenon. The origin of the uncertainty relation can be attributed as
%a mathematical property of Fourier analysis [1] which describes quantum
%mechanics as a wave theory. Recent years have seen effort in establishing
%a stronger relation based on information-theoretical considerations [2][3].
%
In realistic conditions, however, quantum systems are often prepared and
studied
at finite temperatures where thermal fluctuations permeate. At high
temperatures the equipartition theorem of classical statistical mechanics
imparts for each degree of freedom an uncertainty of $kT/2$.
Thus the uncertainty function for a one-dimensional particle
approaches the limit
\be
U_T ^{MB} \approx ({kT\over\Omega})^2
{}~~~(high~T,~~ classical~ statistical~ mechanics),
%                                                               \eqno(2)
\te
where $\hbar \Omega$ is the energy of a normal mode with physical frequency
$\Omega$.
This result, obtained by assuming that the system obeys the
Maxwell-Boltzmann distribution, is  usually regarded as the classical limit.
For a system of bosons in equilibrium at temperature $T$, the application
of canonical ensemble gives the result in quantum statistical mechanics
as
\be
U_T ^{BE} = {\hbar^2 \over 4} [ \coth ({{ \hbar \Omega} \over {2 kT}})]^{2}
{}~~~(all~ T,~~quantum~ statistical~ mechanics),
%                                                                \eqno(3)
\te
which interpolates between the two results (3.1) and (3.2) at $T=0$ and
$T>> \hbar \Omega /k$. This result applies to a system already in
equilibrium at temperature $T$.

Our purpose here is to study the corresponding {\it non-equilibrium} problem.
At time $t_0$ we put the system in contact with a heat bath at temperature
$T$ and follow its time evolution. We want to see how the uncertainty
function $U_T(t)$ changes from the initial quantum fluctuation-dominated
condition to a later thermal fluctuation-dominated condition. By comparing
this result with the decoherence studies reported in Sec. 2
%recently carried out [5] [6],
where two characteristic times---the decoherence time $t_{dec}$ and the
relaxation time $t_{rel}$---are defined, one can apply the physics of
these two processes involved to examine the issues stated above, i.e.,
%\noindent 1) the realization of the basic tenets of quantum statistical
%mechanics from quantum dynamics;
1) the relation between quantum and thermal fluctuations; and
2) their role in  quantum to classical transition.

%For example,
%one can ask when the two postulates of equilibrium statistical mechanics
%are satisfied? When the thermal fluctuation overrides the quantum
%%fluctuations?
%and when the system assumes classical behavior? They can be seen to be
%related to each other and largely determined by the two time scales.

%Quantum statistical mechanics of a macroscopic system is derived from
%the quantum dynamics of its microscopic constituents under two basic
%postulates [7]: i) random phase, and ii) equal {\it a priori} probability.
%The first condition enables one to assign probability distributions to
%a system occupying certain quantum states. It requires the disappearance
%of interference terms in the wave function or that the density matrix
%of the system be approximately diagonal. The second condition ensures
%that the system when put in contact with a large bath equilibrates with
%it. We want to examine the processes by which these two conditions are
%attained from a more primitive level, starting with the microdynamics of
%a system of quantum particles. Specifically, we want to see if there is
%a characteristic time  when the phase information is lost (Postulate i)
%and another time when the system attains equilibrium with its surrounding
%so that all accesible states are equally probable (Postulate ii).

\section{Noise and Decoherence, Fluctuation and Dissipation}
\setcounter{equation}{0}

We use a simple model of a quantum open system to examine these issues.
%We study the uncertainty principle at finite temperature because it is
%a simple problem where these issues manifest clearly.
%The model we use is that of
Consider a collection of coupled harmonic oscillators where one is
distinguished
    as
the system of interest and the rest as bath. We use the influence functional
method to incorporate the statistical effect of the bath on the system.
As the microdynamics is explicit in this approach, one can study how the result
depends on the properties of the bath and the system-bath interaction. This
mode
   l
has been studied extensively, so we will only present the main results.
(See, e.g., \cite{ZhangPhD} for details.)
%One can also see from this calculation the time-development of
%the system from the initial quantum to the final classical regimes.
%One can also stipulate the conditions that the basic hypothesis i) and ii)
%are satisfied, or , more interestingly, when they fail.
%In our studies
%the low temperature weak coupling and supraohmic environment may
%invalidate these postulates, which implies that quantum statistic mechanics
%cannot be applied to the descriptions of such systems.

Our system is a Brownian particle with mass $M$ and natural frequency $\Omega$.
The environment is modeled by a set of $n$ harmonic oscillators with mass $m_n$
and natural frequency $\omega_n$. The particle is coupled linearly to the $n$th
oscillator with strength $C_n$. The action of the combined system and
environmen
   t
is
\be
S[x,q] = S[x]+S_b[q]+S_{int}[x,q]
\te
$$
 = \int\limits_0^tds\Biggl[
    \Bigl\{{1\over 2}M\dot x^2-{1\over 2}M\Omega_0^2 x^2\Bigr\}
  + \sum_n\Bigl\{{1\over 2}m_n\dot q_n^2
  - {1\over 2}m_n\omega^2_nq_n^2 \Bigr\}
  + \sum_n\Bigl\{-C_nxq_n\Bigr\}\Biggr]
$$
\noindent where $x$ and $q_n$ are the coordinates of the particle and the
oscillators respectively, and $\Omega_0$ is the (bare) frequency of the
particle. We are interested in how the environment affects the system in
some averaged manner.  The quantity containing this information is the reduced
density matrix of the system $\rho_r(x,x')$ obtained from the full density
operator of the system and environment $\rho(x,q;x'q')$ by tracing out the
environmental degrees of freedom $ (q,q') $
\be
\rho_r(x,x',t)
=\int\limits_{-\infty}^{+\infty}dq
 \int\limits_{-\infty}^{+\infty}dq'
 \rho(x,q;x',q',t)\delta(q-q')                    %\eqno(5)
\te
\noindent The reduced density matrix evolves under the action of the
propagator \\
 $J_r(x,x',t~|~x_i,x'_i,0) $ in the following way:
\be
\rho_r(x,x',t)
=\int\limits_{-\infty}^{+\infty}dx_i
 \int\limits_{-\infty}^{+\infty}dx'_i~
 J_r(x,x',t~|~x_i,x'_i,0)~\rho_r(x_i,x'_i,0~)   %\eqno(6)
\te
In general, this is a very complicated expression since the evolution
operator $J_r$ depends on the initial state. If we assume that at a given
time $t=0$ the system and the environment are uncorrelated, i.e. that
\be
\hat\rho(t=0)=\hat\rho_s\times\hat\rho_e,       %\eqno(7)
\te
\noindent then the evolution operator for the reduced density matrix can
be written as
\be
J(x_f,x'_f,t~|~x_i,x'_i,0)
=\int\limits_{x_i}^{x_f}Dx
 \int\limits_{x'_i}^{x'_f}Dx'~
 \exp{i\over\hbar}\Bigl\{S[x]-S[x']\Bigr\}~F[x,x']
                                                 %\eqno(8)
\te
\noindent where $F[x,x']$ is the Feynman-Vernon influence functional.
Assuming (4.4) and that the environment is initially in thermal
equilibrium at a temperature $T=\beta^{-1}$
%for the problem described by (4),
the influence functional for (4.1) can be computed exactly.
The result is well known \cite{FeyVer,CalLeg83}:
$$
F[x,x']=\exp\biggl\{
 -{i\over\hbar}\int\limits_0^tds_1\int\limits_0^{s_1}ds_2
   \Bigl[x(s_1)-x'(s_1)\Bigr]\eta(s_1-s_2)
   \Bigl[x(s_2)+x'(s_2)\Bigr]
$$
\be
 -{1\over\hbar}\int\limits_0^tds_1\int\limits_0^{s_1}ds_2
   \Bigl[x(s_1)-x'(s_1)\Bigr]\nu(s_1-s_2)
   \Bigl[x(s_2)-x'(s_2)\Bigr]\biggl\}
\te
\noindent The non-local kernels $ \eta $ and $ \nu $ are the dissipation
and noise kernels defined respectively as
\be
\nu(s) =\int\limits_0^{+\infty}d\omega~I(\omega)
  \coth({{\hbar \omega} \over {2 k T}})  ~  \cos\omega s
\te
and
\be
\eta(s)={d\over ds}~\gamma(s)
\te
where
\be
\gamma(s)=\int\limits_0^{+\infty} d\omega~{I(\omega)\over\omega}~\cos\omega s.
\te
\noindent Here $I(\omega)$ is the spectral density function of the environment,
\be
I(\omega)= \sum\limits_n{\delta(\omega -\omega_n)}
           {{C^2_n}\over{2 m_n \omega_n}}.           %\eqno(11)
\te
An environment is classified as ohmic $I(\omega)\sim\omega$, supra-ohmic
$I(\omega)\sim\omega^n , n>1$ or sub-ohmic $n<1$. The most studied ohmic
case corresponds to an environment which induces a dissipative force
linear in the velocity of the system. An example which we have studied
has spectral density given by \cite{LegRMP}
\be
I(\omega)
=M\gamma_0 \omega ({\omega\over \tilde\omega})^s
e^{-{\omega^2\over\Lambda^2}}                      %\eqno(12)
\te
\noindent where $\tilde\omega$ is a frequency scale usually taken to be
the cut-off frequency $\Lambda$.

%In terms of the kernels $ \eta $ and $\nu$, the propagator for the reduced
%density matrix can be expressed as
%\be
%J_r(x_f,x'_f,t~|~x_i,x'_i,0)
%=\int\limits_{(x_i,x'_i,0)}^{(x_f,x'_f,t)}DxDx'~
% \exp{i\over \hbar} A[x,x']
%                                                    %\eqno(13)
%\te
%\noindent where the influence action is given for this system and environment
%by
%$$
%\eqalign{
%A[x,x']
%&= S[x] - S[x']
%-~2\int\limits_0^tds_1\int\limits_0^{s_1}ds_2~
%  Y(s_1)\eta(s_1-s_2)X(s_2)~+\cr
%&\qquad+i\int\limits_0^t ds_1\int\limits_0^{s_1}ds_2
%  Y(s_1)\nu(s_1-s_2)Y(s_2) \cr }
%                                                    % \eqno(14)
%$$
%\noindent Here we have introduced
%$$
%{\eqalign{
%&X= {x+x'\over 2} \cr
%&Y=x'-x.\cr}}                                       % \eqno(15)
%$$
%\noindent as the ``center of mass'' and ``relative'' coordinates.

%The real and imaginary exponents of $F[x,x']$ are usually regarded as
%responsible for dissipation
%and noise respectively, thus the names dissipation and noise kernels are given
%to $\eta$ and $\nu$.
The most general environment gives rise to nonlocal
dissipation and colored noises. (We refer the reader to Ref \cite{HPZ2}
for a discussion of the generalized fluctuation-dissipation relation
and the time scales for the relevant processes.)
The propagator for a general environment has been calculated before
\cite{HPZ1}:
$$
J(x_f,x'_f,t~|~x_i,x'_i,0)
 = Z_0(t)  \exp{i\over\hbar} \Bigl\{
   \Bigl[\dot u_1(0)X_i+\dot u_2(0)X_f\Bigr]y_i
  -\Bigl[\dot u_1(t)X_i
$$
$$
 +\dot u_2(t)X_f\Bigr]y_f\Bigr\}
 \times\exp{-1\over\hbar}\Bigl\{a_{11}(t)y_i^2
  +[a_{12}(t)+a_{21}(t)]y_iy_f
  +a_{22}(t)y_f^2\Bigr\}                        % \eqno(13)
$$
\noindent where $ X=(x+x')/2 $ and $y=x'-x$, $u_a(s)$ are elementary functions
obtained as solutions of differential equations involving the dissipation
kernel,
and the coefficeints $ a_{ij}(t) $ are obtained from  integrals
involving the noise kernel.

\section{ Uncertainty Principle at Finite Temperature}
\setcounter{equation}{0}

We now consider a Brownian oscillator with an initial wave function
\be
\psi(x_i,0) = N_0~e^{-\frac{x_i^2}{4\sigma^2}}      %\eqno(17)
\te
\noindent where $\sigma$ is the initial spread of the Gaussian packet. One
can calculate $\rho_r(x_f, x'_f, t)$ by performing the Gaussian integrals
over $x_i$ and $x'_i$ and get
\be
\rho(x_f, x'_f, t)
=\Bigl[Z_0(t) N_0^2 {\pi\over\sqrt{det {\bf H}}}\Bigr]
\exp\Bigl\{-{1\over 2} {\bf X^T}{\bf Q}^{-1} {\bf X}\Bigr\}
                                                   %  \eqno(18)
\te
The prefactor (terms within the square bracket) $\tilde N_0 (t)$
depends only on time. Here ${\bf X}=(X,y)^{\bf T}$ and  $Q_{ij}(t)$ is a
$2\times 2$ matrix whose elements are given in \cite{HuZhaUncer}.

To calculate the averages of observables, it is convenient to use the
Wigner function defined as
\be
W(X,p,t) = \int dy e^{{i\over \hbar} p y} \rho (X- {y \over 2},
             X+ {y \over 2}, t),                      %   \eqno(19)
\te
%
%\noindent or its inverse
%
%\be
%\rho (X- {y \over 2}, X+ {y \over 2}, t) = \int {{dp} \over
%{2\pi \hbar}} e^{{-i \over \hbar} p y} W(X, p, t)     %   \eqno(24b)
%\te
%
\noindent The quantum average of an observable, e.g., $x^n$, with
respect to a pure state is given by
\be
< x^n>_0 = \int dx x^n \rho(x,x,t)
         = \int dx \int {{dp} \over {2 \pi \hbar}} x^n W (X, p, t)
                                                        %\eqno(20a)
\te
\noindent and
\be
<p^n>_0 =  \int dx \int {{dp} \over {2 \pi \hbar}} p^n W (X, p, t)
                                                        %\eqno(20b)
\te
\noindent Similar relations exist between $\rho_r$ and $W_r$. The averages
with respect to a mixed state now weighted by  $\rho_r$ or $W_r$ have both
quantum and thermal contributions. We get
\be
<x^2>_T = {1\over Q_{11}(t)}                          % \eqno(21a)
\te
\noindent and
\be
<p^2>_T = \hbar^2 {\det Q~(t)\over Q_{11}(t)}
                                                       % \eqno(21b)
\te
\noindent From them, with $ (\Delta x)_T^2=<x^2>_T-<x>_T^2 $ and
$ (\Delta p)_T^2=<p^2>_T-<p>_T^2 $,
\be
U_T (t) = (\Delta x)_T^2 (\Delta p)_T^2
        = \hbar^2 {\det Q(t) \over [Q_{11}(t)]^2 }.
        = \hbar^2 \Bigl\{ {Q_{22}(t) \over Q_{11}(t)}
        - {Q_{12}(t) \over Q_{11}(t)}
          {Q_{21}(t) \over Q_{11}(t)} \Bigr\}.
                                                        % \eqno(22)
\te
The exact result was obtained in \cite{HuZhaUncer} by solving the
equations for the $u_i(s)$ functions and the $a_{ij}(t)$ coefficients
numerically.

For an ohmic environment
$ \gamma (t) = 2 \gamma_0 \delta (t) $, and $a_{ij}(t)$, $u_i(t)$ are
simple harmonic and exponential functions. Assume a quantum
minimum-uncertainty initial state where ${\hbar\over 2\sigma^2\Omega}=1$
(which is also the ground state of a harmonic oscillator),
we get for weak couplings (small $\gamma_0$) at all temperatures,
\be
U_T(t)={\hbar^2\over 4}\Bigl[
   e^{-\gamma_0 t} + \coth({\hbar\Omega\over 2kT})
   (1- e^{-\gamma_0 t})\Bigr]^2.
                                                  % \eqno(24)
\te
\noindent This is a simple, clean and intuitively clear result. We see that
there are two factors at play here: time and temperature. Time is measured
in units of the relaxation time proportional to $t_{rel}=\gamma_0^{-1}$, and
temperature is measured with reference to the ground state energy
$\hbar \Omega /2$ of the system. At $t=0$, when the initial uncorrelated
conditions (4.4) is assumed valid, $ U_T(0)=\hbar^2/4$,
which is the Heisenberg relation (3.1). At $t \approx \gamma_0^{-1}$
the system begins to equilibrate with the bath. At very
long time ($t>>\gamma_0^{-1}$), $ U_T(t) $ approaches $ U^{BE}_T $ [(3.3)]
at finite temperature, or $ U^{BM}_T $ [(3.2)] at high temperature.
That means the system (the Brownian particle) approaches an equilibrium
quantum statistical system.

Now call $z= {\hbar \Omega\over 2kT}$. At zero temperature, $\coth z=1$ and
$U_T(t)= U_0^{QM}$ as in (3.1) at all times, as expected. At high temperature
($\coth z \approx 1/z$) and at short times ($t<<\gamma_0^{-1}$) this simplifies
to
\be
U_T(t) = {\hbar^2\over 4}
\Bigl[ 1+\Bigl({2kT\over \hbar\Omega}-1\Bigr)\gamma_0t
       + O(t^2) \Bigr]                                 %  \eqno(25)
\te
For a finite temperature ohmic bath (with weak coupling),
there always exist a  time  ($t> \gamma_0^{-1}$ and
$e^{-\gamma_0t} << coth z$ ) such that
\be
U_T(t) = (kT/\Omega_0)^2                               %  \eqno(26)
\te
which is the classical limit $U_T^{MB}$ [(3.2)].

These simple  expressions are revealing in several aspects: Note that in
the expression for short time behavior (5.10) the first term is the ubiquitous
quantum fluctuation contribution,
the second term is the thermal contribution, which depends
on the initial spread and increases with increasing dissipation and
temperature.

The time when thermal fluctuations overtake quantum fluctuations is when the
second term in the square bracket becomes larger than unity which occurs at
(the temperature is higher than the ground state energy by assumption)
\be
t_1 = {{ \hbar \Omega_0} \over {2 \gamma_0 kT}}    % \eqno(26)
\te
\noindent This is indeed  the decoherence time scale $t_{dec}$ \cite{Zurek},
the time when the off-diagonal components of the reduced density matrix
diminishes to zero.
%and the first postulate of quantum statistical mechanics becomes valid.
The second time scale is the relaxation time scale,
$t_{rel} = \gamma_0^{-1}$ , when the particle reaches
equilibrium with the environment.
%It is at this time that the second postulate
%of quantum statistical mechanics becomes valid and equilibrium QSM can be
%applied to the combined system + environment.
After this, for ohmic and subohmic
environments the uncertainty relation takes on the Bose-Einstein form (3.3).
At high temperatures the system reaches the Maxwell-Boltzmann limit and the
uncertainty relation takes on the classical form (3.2). For weak coupling and
supraohmic environments at low temperature, the highly nonlocal frequency
respon
   se
makes it difficult for the system to settle down. The decoherence time scale is
longer, and the relaxation can even be incomplete.
%
%[See Fig. 1, case C, where the asymptoptic value of $U_T(t)$ is lower than
%that of (3)].
%
This is the regime where one expects to find more intricate
behavior in the interplay of quantum and thermal effects.
%The nonohmic results
%and details of the present study are to be presented elsewhere \cite{HRZ}.

%This should, however, not be viewed as the
%threshhold of classical domain, in the decoherence sense. It is the limit
%when classical kinetic theory, specifically, the Maxwell-Boltzmann
%distribution becomes applicable to a system already in equilibrium with
%a heat bath.
%This study also corrects a common conception which associates thermal with
%classical. As we can see, thermal in a general sense can better be understood
%as the statistical average over a large number of degrees of freedom as in
%the notion of a quantum open system.
%
%For the more general nonohmic environment, the behavior is more complex.
%For low temperatures or supraohmic environments, quantum behavior is
%expected to persist longer, and the interplay of quantum and thermal
%fluctuations more intricate.

% \newpage
%\vskip 0.2in

\section{Chaos in Quantum Open Systems}

In this talk we have only partially addressed
half of the theme of this conference,
i.e., the relation of quantum and classical.
The other part on nonintegrability is not touched upon, because we don't know
enough about this important and fascinating subject. In light of what
we have discussed above, i.e., viewing classical behavior as the
result of decoherence on a quantum system,
it is nonetheless interesting to raise a few general questions, naive as they
may be to the experts:

A direct question to ask is nonintegrability in both quantum and classical
open systems: i.e.,
How would the introduction of an environment alter the intrinsically
chaotic or nonchaotic character in the dynamics of a system? Is chaotic
behavior likely to be enhanced or quenched? Some aspects of this problem
have been studies as chaos in systems with noise \cite{ChaosNoise}.
Usually if the system's nonlinear dynamics shows chaotic behavior,
the presence of an environment which adds a stochastic force is expected
to modify the chaotic behavior only quantitatively. If the coupling
of the system with the environment is strong and nonlinear, qualitative
changes in the system's dynamics are likely.
%and growth \cite{Stanley}. The emergence of persistent
%structures in the classical world from quantum dissipative systems is another
%fundamental problem
%%The effect of noise and
%dissipation on a quantum integrable system is an open problem.
Although structures and forms are usually discussed in the context of classical
dissipative systems,
it would be very gratifying if one can understand the rules from which
such macroscopic structures \cite{Stanley} in the classical world
emerge from the microscopic dynamics of a quantum system
\cite{GelHar2,Woo,BalVen,HuSpain}.
Nonlinear dissipative dynamics in a quantum open system may
play an important role in this issue.

An indirect, and perhaps conceptually more subtle problem is the following:
If we view classical behavior as a macroscopic phenomenon
emergent from a quantum open system, will it provide us with some
insight to resolve the apparent puzzle that there are chaotic behavior in the
classical equation of motion
but usually not in the associated quantum equation of motion?
Is it easier to account for nonintegrability in an effective theory than a
fundamental theory? In the viewpoint we presented above, while
a quantum closed
system is completely describable by a fundamental theory, vis., quantum
mechanics, the corresponding classical theory is only an effective theory
in the sense that it has incorporated the backreaction effect of the
environment
   ,
which is subjected to coarse-graining approximations.
(In such a process usually a simple dynamical equation is transformed
into an integro-differential equation of motion).
Thus classical dynamics is a limiting form of an effective quantum theory.
One can study how under these statistical procedures
(coarse-graining, decoherence and backreaction)  an integrable
quantum system becomes a classical chaotic system. Conversely,
a more difficult problem is to
take a classical model which exhibits chaotic behavior and determine
if it can be obtained through certain reduction schemes
as the macroscopic limit of an associated set
of integrable quantum models.
This is, of course,  a many-to-one relation,
depending on the choice of schemes. The nonintegrability of the classical
system
could then be seen as an outcome both of the original quantum
dynamics and the quantum to classical reduction scheme.\\

%\footnote{A simple and practical answer to this question could be that
%the classical systems one ordinarily studies are mostly
%underdamped systems, i.e.,in the regime of small or neglible dissipation,
%where noise and fluctuations have little effect on the nonlinear dynamics
%in these systems which give rise to chaos.}
%The system should be nonlinear for one to see the effect of
%the environment on the nonintegrability.}
%linear coupling model studied here is not sufficient to address these
%%problems.
%A more intricate problem is to see a nonintegrable system
%behaves at the threshold of openess.

\section{Acknowledgements}

We thank Esteban Calzetta, Salman Habib, Juan Pablo Paz and Sukanya Sinha
for many lively discussions on basic issues of quantum mechanics and
quantum cosmology.
This work is supported in part by the National Science Foundation under
grant PHY 91-19726.

\end{document}